\documentclass[hyper,letterpaper,12pt]{article}
\usepackage{a4wide}
\usepackage{amsmath}
\usepackage[
      colorlinks=true,
      linkcolor=blue,
      urlcolor=blue,
      filecolor=black,
      citecolor=blue,
            ]{hyperref}
\usepackage{color}
\usepackage{graphicx}
\usepackage{amsfonts}
\usepackage{amssymb}
\usepackage{epstopdf}

\def\beq{\begin{equation}}
\def\eeq{\end{equation}}

\begin{document}
\title{\bf \Large  Entanglement Entropy in Holographic P-Wave Superconductor/Insulator Model}

\author{\large
~Rong-Gen Cai$^1$\footnote{E-mail: cairg@itp.ac.cn}~,
~~Li Li$^1$\footnote{E-mail: liliphy@itp.ac.cn}~,
~~Li-Fang Li$^1$$^3$\footnote{E-mail: lilf@itp.ac.cn}~,
~Ru-Keng Su$^2$\footnote{Email: rksu@fudan.ac.cn}\\
\\
\small $^1$State Key Laboratory of Theoretical Physics,\\
\small Institute of Theoretical Physics, Chinese Academy of Sciences,\\
\small Beijing 100190,  China.\\
\small $^2$ Department of Physics, Fudan University,  Shanghai 200433, China \\
\small $^3$State Key Laboratory of Space Weather, \\
\small Center for Space Science and Applied Research, Chinese Academy of Sciences,\\ 
\small Beijing 100190, China.}
\date{\today}
\maketitle

\begin{abstract}
\normalsize We continue our study of entanglement entropy in the holographic superconducting phase transitions. In this paper we consider the holographic p-wave superconductor/insulator model, where as the back reaction increases, the transition is changed from second order to  first order. We find that unlike the s-wave case, there is no additional first order transition in the superconducting phase. We calculate the entanglement entropy for two strip geometries. One is parallel to the super current, and the other is orthogonal to the super current. In both cases, we find that the entanglement entropy  monotonically increases with respect to the chemical potential.
\end{abstract}

\tableofcontents

\section{ Introduction}

The entanglement entropy nowadays plays an important role in understanding some characterizations in quantum field theory and many-body physics (see, for example, Refs.~\cite{2006PhRvB..73x5115R,Amico:2007ag,Eisert:2008ur}). The entanglement entropy, as the name suggests, measures how a given quantum system is entangled or strongly correlated. Further, it can also be used to distinguish different phases and corresponding phase transitions and is considered as a useful tool for keeping track of the degrees of freedom of strongly coupled systems. For a given system, the entanglement entropy of one subsystem with its complement is defined as the von Neumann entropy. However, the calculation of entanglement entropy is found to be very difficult except for the case in $1+1$ dimensions.

In the framework of AdS/CFT correspondence~\cite{Maldacena:1997re,Gubser:1998bc,Witten:1998qj}, a holographic method to calculate the entanglement entropy has been proposed in Ref.~\cite{Ryu:2006bv}. Following Ref.~\cite{Ryu:2006bv},
for a conformal field theory  (CFT) which has a dual gravitational configuration  living in one higher dimension, the entanglement entropy of the CFT in a subsystem $\mathcal{A}$ with its complement is given by finding the minimal area surface $\gamma_\mathcal{A}$ extended into the bulk with the same boundary $\partial\mathcal{A}$ of $\mathcal{A}$ (see Refs.~\cite{Nishioka:2009un,Takayanagi:2012kg} for reviews). Namely, the entanglement entropy of $\mathcal{A}$ with its complement is given by the ``area law"
\begin{equation}\label{law}
S_\mathcal{A}=\frac{2\pi}{\kappa^2} Area(\gamma_\mathcal{A}),
\end{equation}
where $\kappa^2\equiv 8\pi G $ is related to the gravitational constant in the bulk.

Holographic superconductor model was first constructed in Refs.~\cite{Hartnoll:2008vx,Hartnoll:2008kx} where the model is a s-wave one since the condensed field is a scalar field dual to a scalar operator in the field theory side. The holographic approach was also generalized to the p-wave case~\cite{Gubser:2008wv} and d-wave case~\cite{Chen:2010mk,Benini:2010pr}. It is interesting to note that the back reaction of matter field on the background geometry will change the order of the holographic p-wave superconductor phase transition from second order to first order~\cite{Ammon:2009xh}. On the other hand, the effect of higher derivatives on the holographic p-wave model was studied in Refs.~\cite{CNZ1,CNZ2,Momeni:2012ab}. And near critical point, an analytic study on the holographic p-wave superconductor/conductor phase transition has been made in the probe limit~\cite{Zeng:2010zn,new1}. In Ref.~\cite{Nishioka:2009zj}, a holographic s-wave superconductor/insulator phase transition model was built at zero temperature. Combining the conductor/superconductor phase transition with the insulator/superconductor phase transition, Horowitz and Way~\cite{Horowitz:2010jq} described a complete phase diagram for a holographic s-wave superconductor/conductor/insulator system. In addition, the holographic p-wave superconductor/insulator phase transition was discussed in Ref.~\cite{Akhavan:2010bf}, and Ref.~\cite{Cai:2011ky} carried out an analytic study for the holographic superconductor/insulator phase transition. While various properties of holographic superconductor models have been intensively investigated in the literature, in this paper we are interested in the behavior of entanglement entropy in the holographic superconductor models. Refs.~\cite{Albash:2012pd,Cai:2012nm,Arias:2012py} studied the behavior of entanglement entropy in holographic conductor/superconductor phase transition models, including s-wave and p-wave cases. It showed that the entanglement entropy decreases as one lowers temperature, indicating the degrees of freedom are reduced at lower temperature, and that the entanglement entropy can tell us not only the occurrence of the phase transition, but also the order of the phase transition. It indicates that entanglement entropy is indeed a good probe to phase transition. In Refs.~\cite{Cai:2012sk,Cai:2012es} we investigated the behavior of entanglement entropy in the holographic s-wave superconductor/insulator model, and found that the entanglement entropy as a function of chemical potential is not monotonic in the superconducting phase: at the beginning of the transition, the entropy first increases and arrives at its maximum at some chemical potential, and then decreases monotonically.

In this paper we continue our study of the entanglement entropy in holographic superconductor phase transitions. We here focus on the holographic p-wave superconductor/insulator model, paying main attention on whether those properties of entanglement entropy found in the previous studies are universal or not, in particular, on the non-monotonic behavior of the entropy in the superconductor/insulator phase transition. This will be helpful to further understand the superconductor/insulator transition. Furthermore, there are also other motivations. It is well known that there is no thermal phase transition in $(2+1)$-dimensional field theories due to the Mermin-Wagner theorem or Coleman theorem, which states that in quantum field theory,
continuous symmetries cannot be spontaneously broken at finite temperature in systems with sufficiently short-range interactions in spatial dimensions $d \le 2$, which at least tells us that the phase transition can be affected by the space-time dimension of the system. However, one indeed observes the superconducting phase transition in the gravity dual of $(3+1)$-dimensional AdS black hole backgrounds~\cite{Hartnoll:2008vx}. This might be caused by the suppression of the large fluctuations in the large $N$ limit, which is supposed to be one of conditions for the validness of the AdS/CFT correspondence.  In the framework of AdS/CFT correspondence, we can study the strongly coupled systems simply through their dual weakly interacting gravity theories. So a natural question arises that how the order of the phase transition of the strongly coupled systems can be affected by the dimension of space-time. Indeed, it is observed in Ref.~\cite{Arias:2012py} that in contrast to the $(3+1)$-dimensional holographic p-wave superconductor/conductor phase transition, the phase transition in the $(2+1)$-dimensional case is always second order, regardless of the strength of the back reaction. To further stress the issue and enrich our understanding of the holographic superconductor phase transitions, therefore it would be helpful to study  the full back reacted holographic p-wave superconductor/insulator model.

The p-wave superconducting phase is characterized by the condensation of the vector ``hair"~\cite{Gubser:2008zu,Gubser:2008wv}, while the normal insulator phase is described by a pure AdS soliton solution. As one increases the chemical potential, the pure AdS soliton background will become unstable to develop the vector ``hair". The emergence of  the ``hair" induces the symmetry breaking and gives a finite vacuum expectation value of the dual current operator in the field theory side, which plays the role of order parameter in the holographic phase transition. We will choose the condensate of the current operator along one spatial direction. Thus, the $U(1)$ symmetry as well as the spatial rotational symmetry are broken in the superconducting phase. The order of the transition from the insulator phase to superconducting phase can be either second order or first order, depending on the strength of the back reaction of matter field on the bulk geometry. For the holographic s-wave superconductor/insulator case, one can find a new first order phase transition inside the superconducting phase~\cite{Horowitz:2010jq,Cai:2012es}. We will see that this does not happen in the p-wave model. In view of the anisotropy in the superconducting phase, we will consider two strip geometries with a finite width along directions parallel to the super current and orthogonal to the super current, respectively. The value of entanglement entropy calculated in former setup is larger than the one in the latter case for given parameters. In addition, no matter of the order of the transition, we find the behavior of the entanglement entropy as a function of chemical potential is monotonic. More precisely, the entanglement entropy always increases as the increase of chemical potential. Such result is quite different from the one in the s-wave case.

This paper is organized as follows. In section~\ref{sect:Pwave}, we introduce the holographic model and study the thermodynamics of the transition. Since the system is anisotropic in the superconducting phase, we calculate, in section~\ref{sect:Phee}, the entanglement entropy for strip geometry along spatial direction $x$ and $y$, respectively. The conclusion and some discussions are included in section~\ref{sect:conclusion}. Some numerical details are given in appendix~\ref{sect:details}.


\section{Holographic P-wave superconductor/insulator model}
\label{sect:Pwave}

\subsection{Gravity background}
\label{sect:Pgravity}

Let's begin with the $(4+1)$-dimensional $SU(2)$ Einstein-Yang-Mills theory with a
negative cosmological constant~\cite{Gubser:2008wv}
\begin{equation}\label{action}
S =\int d^5 x
\sqrt{-g}[\frac{1}{2\kappa^2}(\mathcal{R}+\frac{12}{L^2})-\frac{1}{4\hat{g}^2}
F^a_{\mu\nu} F^{a\mu \nu}],
\end{equation}
where $\hat{g}$ is the Yang-Mills coupling constant and $L$ is the AdS radius. The SU(2)
Yang-Mills field strength is given by~\footnote{$\mu,\nu=(t,r,x,y,z)$ denote the indices of space-time and
$a,b,c=(1,2,3)$ are the indices of the SU(2) group generators
$\tau^a=\sigma^a/2i$ where $\sigma^a$ are Pauli matrices.}
\begin{equation}
 F^a_{\mu\nu}=\partial_\mu A^a_\nu-\partial_\nu A^a_\mu + \epsilon^{abc}A^b_\mu
 A^c_\nu,
\end{equation}
where $\epsilon^{abc}$ is the totally antisymmetric tensor with $\epsilon^{123}=+1$. The gauge field is given by
$A=A^a_{\mu}\tau^adx^{\mu}$. Here we define a parameter $\alpha\equiv\kappa/\hat{g}$ which measures the strength of the back
reaction of the Yang-Mills field on the background geometry.

Our ansatz for the metric and Yang-Mills fields are given by~\footnote{ Our ansatz adopted here is very similar with the one in~Ref.\cite{Akhavan:2010bf}. In the case $\alpha\rightarrow0$, the probe limit can work very well. Therefore one can recover the results obtained in the probe limit~\cite{Akhavan:2010bf,Cai:2011ky} when one takes $\alpha =0$.}
\begin{equation}\label{metric}
d s^2 =\frac{L^2}{r^2}\frac{d r^2}{g(r)} + r^2(-f(r)d t^2+h(r)d x^2+d y^2+g(r)e^{-\chi(r)}d \eta^2),
\end{equation}
\begin{equation}\label{gauge}
A=\phi(r)\tau^3 dt+ w(r)\tau^1 dx.
\end{equation}
$g(r)$ vanishes at the tip $r=r_0$ of the soliton. Further, in order to obtain a smooth geometry at the tip $r_0$, $\eta$ should be made with an identification
\begin{equation}
\eta\sim\eta+\Gamma,\qquad \Gamma=\frac{4\pi L e^{\frac{\chi(r_0)}{2}}}{r_0^2 g'(r_0)}\;.
\end{equation}
This gives a dual picture of the boundary theory with a mass gap, which is reminiscent of an insulator phase~\cite{Nishioka:2009zj}.

The independent equations of motion in terms of the above ansatz can be written as follows:
\begin{equation}\label{eoms}
\begin{split}
\phi''+(\frac{3}{r}-\frac{f'}{2f}+\frac{g'}{g}+\frac{h'}{2h}-\frac{\chi'}{2})\phi'-\frac{L^2\omega^2\phi}{r^4 gh}=0, \\
\omega''+(\frac{3}{r}+\frac{f'}{2f}+\frac{g'}{g}-\frac{h'}{2h}-\frac{\chi'}{2})\omega'+\frac{L^2\phi^2\omega}{r^4 fg}=0, \\
f'h'+(\frac{3}{r}+\frac{g'}{g}-\chi')(fh)'-\frac{3fh\chi'}{r}-\frac{4L^2\alpha^2\phi^2\omega^2}{r^6 g}=0,\\
f''+(\frac{5}{r}-\frac{f'}{2f}+\frac{g'}{g}+\frac{h'}{2h}-\frac{\chi'}{2})f'-\frac{2\alpha^2\phi'^2}{r^2}-\frac{2L^2\alpha^2\phi^2\omega^2}{r^6 gh}=0 ,\\
h''+(\frac{5}{r}+\frac{f'}{2f}+\frac{g'}{g}-\frac{h'}{2h}-\frac{\chi'}{2})h'+\frac{2\alpha^2\omega'^2}{r^2}-\frac{2L^2\alpha^2\phi^2\omega^2}{r^6 gf}=0,\\
\chi'-\frac{f'}{f}-\frac{2g'}{g}-\frac{h'}{h}+\frac{8}{r g}-\frac{8}{r}-\frac{2\alpha^2}{3 r}(\frac{\phi'^2}{f}-\frac{\omega'^2}{h})-\frac{2L^2\alpha^2\phi^2\omega^2}{3r^5 fgh}=0,
\end{split}
\end{equation}
where the prime denotes the derivative with respect to $r$ and the tip $r=r_0$ is determined by the condition $g(r_0)=0$.

In order to match the asymptotical AdS boundary, the matter and metric fields near the AdS boundary $r\rightarrow\infty$ should behave as
\begin{equation} \label{boundary}
\begin{split}
\phi&=\phi_0-\frac{\phi_2}{r^2}+\ldots,\quad \omega=\omega_0+\frac{\omega_2}{r^2}+\ldots, \\
\quad f=1+\frac{f_4}{r^4}+\ldots,\quad g&=1+\frac{g_4}{r^4}+\ldots,\quad h=1+\frac{h_4}{r^4}+\ldots,\quad \chi=\frac{\chi_4}{r^4}+\ldots,
\end{split}
\end{equation}
where $\omega_0$ is regarded as the source of the dual current operator. To break the $U(1)$ gauge symmetry spontaneously, we impose $\omega_0=0$. In addition, we choose the condensate along the $x$ direction. Thus, the rotational symmetry in $x-y$ plane is also destroyed in superconducting phase. According to the AdS/CFT dictionary, we can obtain the current condensate $\langle\hat{J}^x_1\rangle=\frac{2\alpha^2}{\kappa^2 L}w_2$, chemical potential $\mu=\phi_0$ and charge density $\rho=\frac{2\alpha^2}{\kappa^2 L}\phi_2$ by reading off the coefficients $w_2$, $\phi_0$ and $\phi_2$ in~\eqref{boundary}, respectively.  Furthermore,  here $f_4$, $g_4$, $h_4$ and $\chi_4$ are all constants.  Let us note that there exists an analytic solution for the equations of motion \eqref{eoms}, which is just the so-called AdS soliton solution
\begin{equation}\label{metric}
\phi(r)=\mu,\quad \omega(r)=\chi(r)=0,\quad f(r)=h(r)=1,\quad g(r)=1-\frac{r_0^4}{r^4}.
\end{equation}
Here $\mu$ is an arbitrary constant, but less than its critical value, which will be showed shortly,   and $r=r_0$ stands for the tip of the soliton. The AdS soliton solution has vanishing $\omega$, thus corresponds to the normal insulator phase. In order to compare physical quantities obtained from different solutions, the boundary geometry must be the same,  in particular, the $\eta$ coordinate must have the same compactification length $\Gamma$. Therefore, we scale all of $\Gamma$ for each solution to be $\pi L$. We work in unites where $L=1$ hereafter, so when $L$ appears in later sections it is set to be one. Numerical details for solving the equations of motion (\ref{eoms}) can be found in appendix~\eqref{sect:details}.
\begin{figure}[h]
\centering
\includegraphics[scale=0.9]{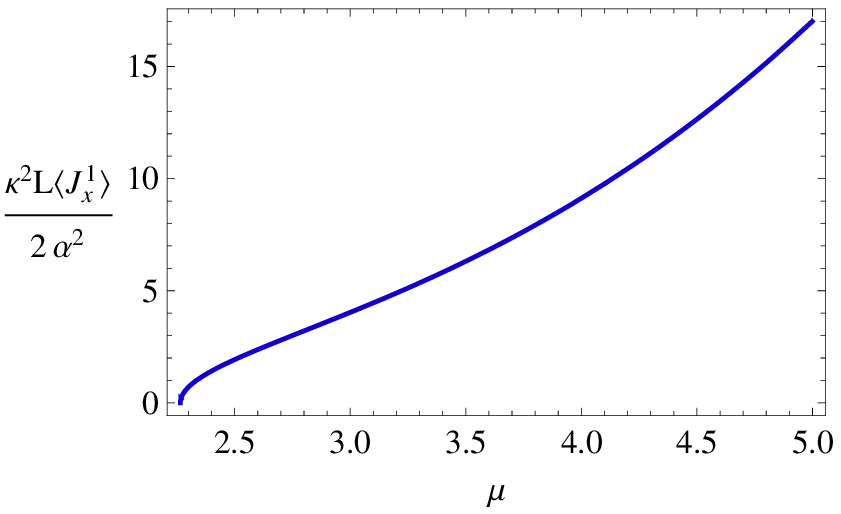}\ \ \ \
\includegraphics[scale=0.91]{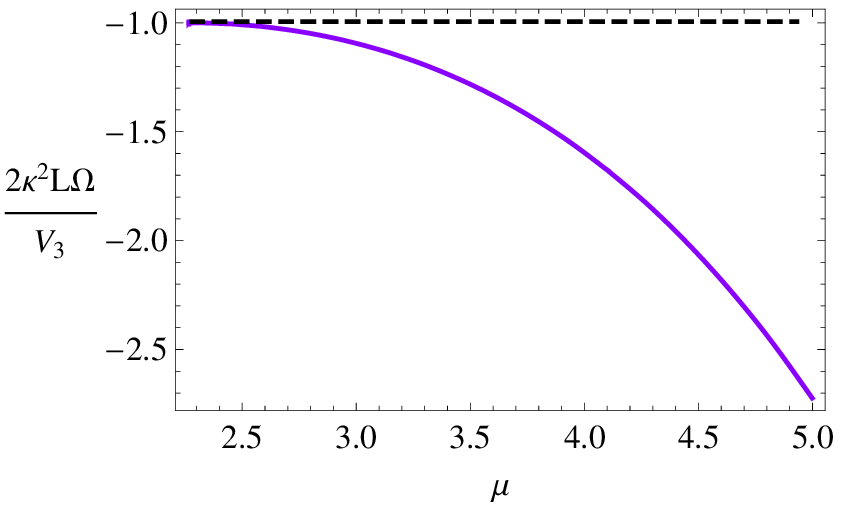} \caption{\label{pphase1} The condensate of current operator
 $\langle\hat{J}^x_1\rangle$ (left plot) versus chemical potential $\mu$ for $\alpha=0.2$. $\Gamma$ is scaled to be $\pi L$. The critical chemical potential in this case is $\mu_c\simeq2.265$. It is a typical second order phase transition which can be seen clearly from right plot, which presents the grand potential of the soliton with vector ``hair" (solid purple) and of the soliton without vector ``hair" (dashed black), with respect to chemical potential.}
\end{figure}

\subsection{Thermodynamics and phase transition}
\label{sect:Pthermo}

To find the thermodynamically favored phases and to determine the critical point of phase transition, we should calculate the grand  potential function $\Omega$ of the system since we are working in grand canonical ensemble in this paper. In gauge/gravity duality the grand potential $\Omega$ of the boundary thermal state is identified with temperature times the on-shell bulk
action in Euclidean signature. The Euclidean action must include the Gibbons-Hawking boundary term for a well-defined Dirichlet
variational principle and further a surface counterterm for removing divergence
\begin{equation}
S_{Euclidean}=-\int d^5 x
\sqrt{g}[\frac{1}{2\kappa^2}(\mathcal{R}+\frac{12}{L^2})-\frac{1}{4\hat{g}^2}
F^a_{\mu\nu}
F^{a\mu\nu}]+\frac{1}{2\kappa^2}\int_{r\rightarrow\infty} d^4x
\sqrt{h}(-2\mathcal{K}+\frac{6}{L}),
\end{equation}
where $h$ is the induced metric on the boundary, and $\mathcal{K}$ is the trace of the extrinsic
curvature. By using of the equations of motion~\eqref{eoms} and the asymptotical expansion of matter and metric functions near the boundary, the grand potential $\Omega$ turns out to be~\footnote{The soliton background has no horizon and the associated Hawking temperature vanishes. But one can introduce an arbitrary inverse temperature $1/T$ as the period of the Euclidean time coordinate. Since the soliton solution is static, the integration over the Euclidean time in the Euclidean action just gives the factor $1/T$, which cancels the temperature factor in the grand potential and leads to a finite grand potential.}
\begin{equation}\label{grand1}
\frac{2L\kappa^2\Omega}{V_3}=g_4,
\end{equation}
where $V_3=\int dx dy d\eta$. Since we have scaled $\Gamma$ to be $\pi L$, $g_4=-1$ for the pure AdS soliton solution, namely, in the normal insulator  phase.

\begin{figure}[h]
\centering
\includegraphics[scale=0.9]{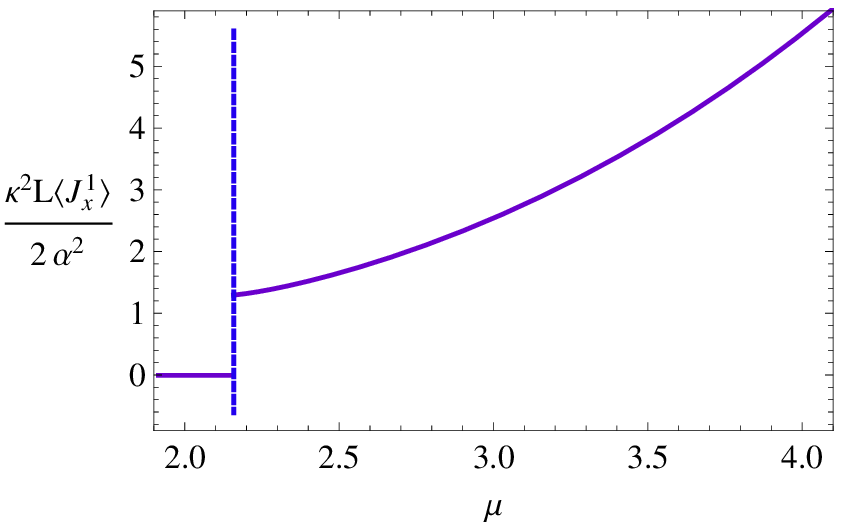}\ \ \ \
\includegraphics[scale=0.91]{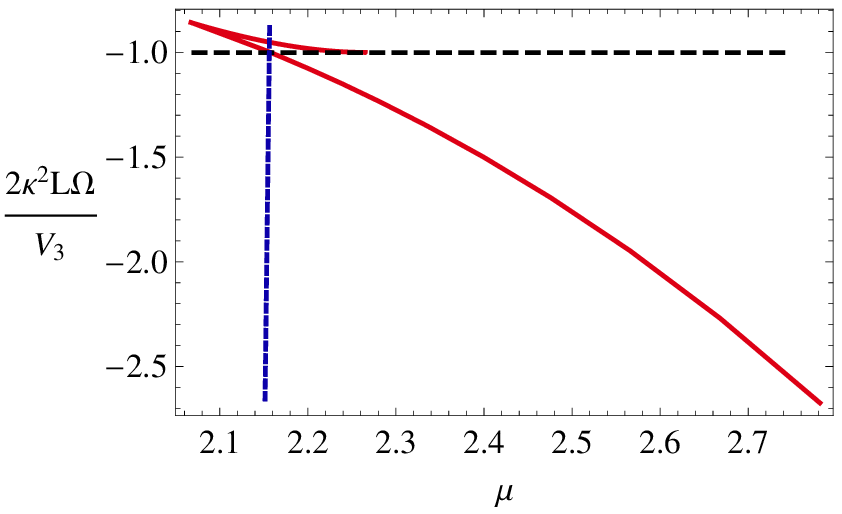} \caption{\label{pphase2} The condensate of current operator
 $\langle\hat{J}^x_1\rangle$ (left plot) versus chemical potential $\mu$ for $\alpha=0.6$. $\Gamma$ is scaled to be $\pi L$.
 The right plot shows the grand potential of the soliton with vector ``hair" (solid red) and the soliton without vector ``hair" (dashed black) with respect to chemical potential. It is a typical first order phase transition and the critical chemical potential denoted by vertical dotted line is $\mu_c\simeq2.157$.}
\end{figure}

\begin{figure}[h]
\centering
\includegraphics[scale=0.9]{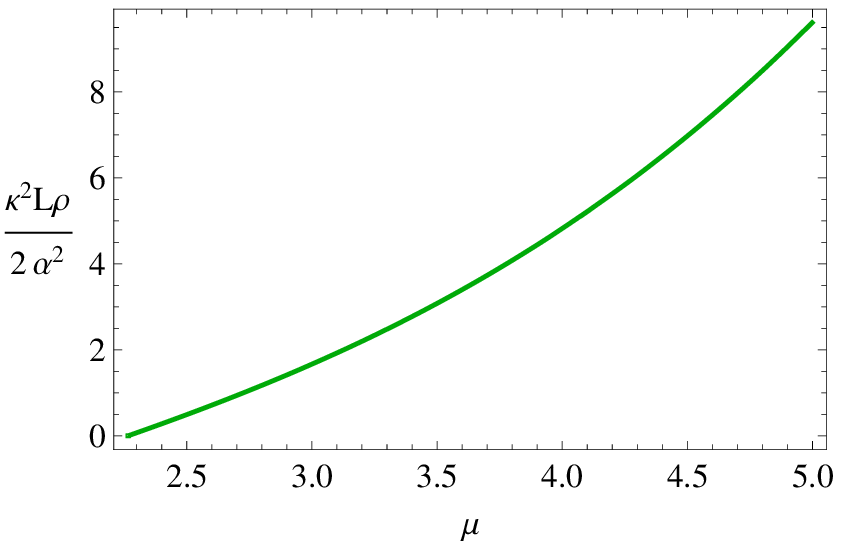}\ \ \ \
\includegraphics[scale=0.9]{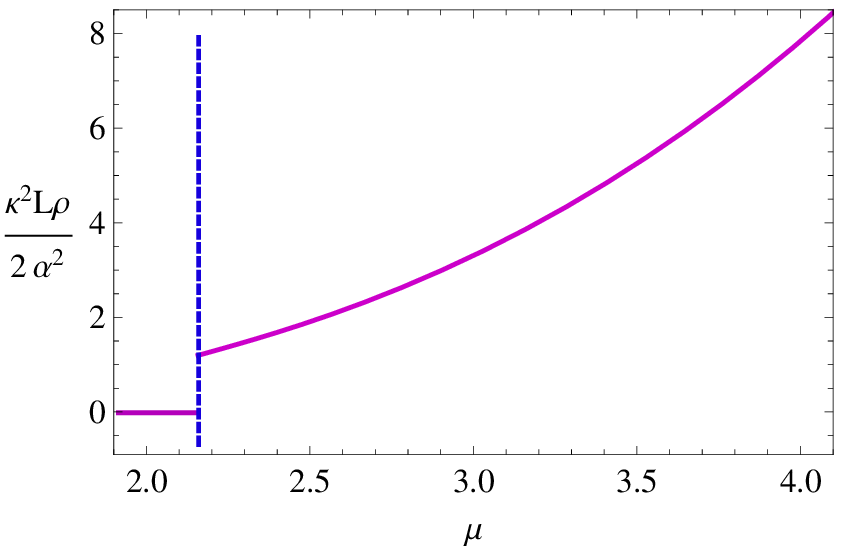} \caption{\label{pcharge} The charge density as a function of chemical potential for $\alpha=0.2$ with $\mu_c\simeq2.265$  (left plot) and $\alpha=0.6$ (right plot), respectively. $\Gamma$ is scaled to be $\pi L$. The critical chemical potential denoted by vertical dotted line is $\mu_c\simeq2.157$ for the right plot.}
\end{figure}

The condensation of vector operator $\hat{J}^x_1$ and the grand potential as a function of chemical potential for $\alpha=0.2$ and $\alpha=0.6$, as two typical examples,  are presented in Figure~\eqref{pphase1} and Figure~\eqref{pphase2}, respectively. Above the critical chemical potential $\mu_c$, the solution with non-vanishing vector ``hair" is thermodynamically favored over the pure AdS soliton solution. The emergency of the vector ``hair" breaks the $U(1)$ gauge symmetry as well as rotational symmetry and represents the superconducting phase. In contrast to the case in Figure~\eqref{pphase1} where $\langle\hat{J}^x_1\rangle$ rises continuously from zero at $\mu_c$, the case in Figure~\eqref{pphase2} has a jump in the $\langle\hat{J}^x_1\rangle$ at $\mu_c$ indicating the first order of the transition. The corresponding charge density is drawn in Figure~\eqref{pcharge}, which behaves qualitatively similar as the condensation. The order of the phase transition can be seen more clearly from the grand potential $\Omega$ with respect to chemical potential. For the first case, the grand potential decreases monotonically as one increases the chemical potential, indicating a continuous phase transition. For the second case, the grand potential with respect to chemical potential develops a ``swallow tail", which is a typical feature in first order transition.
 Thus, as expected, we find that the order of the phase transition depends on the parameter $\alpha$, i.e., the strength of the back reaction. The critical value is $\alpha_c\simeq0.538\pm0.002$,  beyond which the order of the phase transition changes from second order to first order. However, unlike the s-wave case, there is no additional first order transition occurring in the superconducting phase~\cite{Horowitz:2010jq,Cai:2012es}.

In addition, it is worth pointing out that except for the pure AdS soliton solution (insulator phase) and hairy soliton solution (superconducting phase) mentioned above, the action~\eqref{action} admits another two solutions: the black hole solution without hair (conductor phase) and black hole solution with vector hair (another superconducting phase).  A rough phase diagram has been drawn in~Ref.\cite{Akhavan:2010bf}. The details depend on the strength of the back reaction. The dependence of the order of phase transition on the back reaction is mentioned there. In this paper we just focus on the case with zero temperature insultor/superconductor phase transition. In this case, the black hole phase does not appear. In particular, in the zero temperature case, for any values of chemical potential beyond the critical one, the superconducting phase always exists.

\section{Entanglement entropy}
\label{sect:Phee}
Let us begin to study the behavior of entanglement entropy in this p-wave model. Since the system is anisotropic in the superconducting phase, we will consider a straight belt geometry with a finite width $\ell$ along spatial direction $x$ and $y$, respectively.
\subsection{Strip with finite width along $x$ direction}
\label{sect:Pgravity}
We first consider a straight belt geometry with a finite width $\ell$ along the $x$ direction. More specifically, the subsystem $\mathcal{A}$ sits on the slice $r=\frac{1}{\epsilon}$ where $\epsilon\rightarrow0$ is the UV cutoff, and extends in $y$ and $\eta$ directions. The holographic dual surface $\gamma_A$ is defined as a three-dimensional surface
\begin{equation}\label{embed}
t=0,\ \ r=r(x),\ \ -\frac{R}{2}<y<\frac{R}{2}\ (R\rightarrow\infty),\ \ 0\leq\eta\leq\Gamma.
\end{equation}
$R$ is the regularized length in $y$ direction. For this boundary subsystem, there exist two kinds of dual surfaces in the bulk. One is called connected configuration and
the other is disconnected configuration. We first focus on the connected configuration which is smooth at the turning point $r=r_*$: The holographic surface $\gamma_A$ starts from $x=\frac{\ell}{2}$ at $r=\frac{1}{\epsilon}$, extends into the bulk until it reaches $r=r_*$, then returns back to the AdS boundary $r=\frac{1}{\epsilon}$ at $x=-\frac{\ell}{2}$.
\begin{figure}[h]
\centering
\includegraphics[scale=1.3]{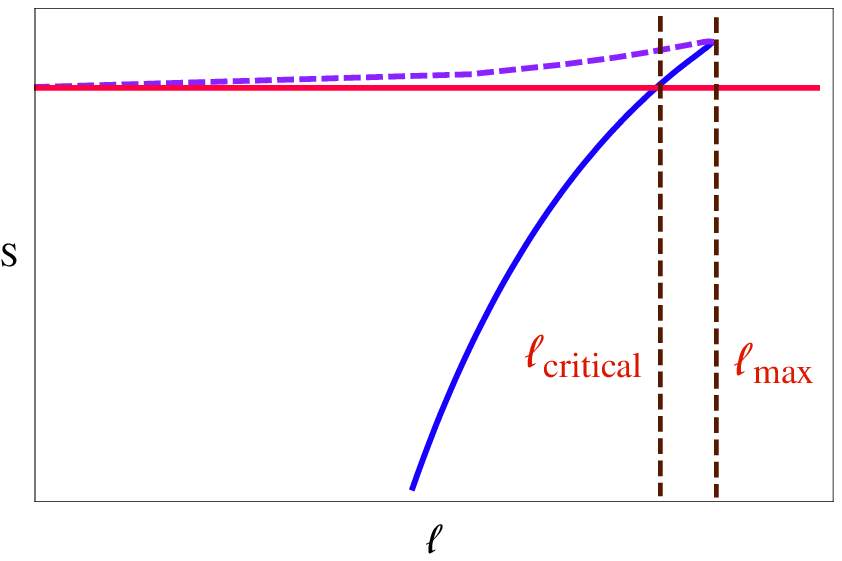}
\caption{\label{entropywidth} The typical configuration for the entanglement entropy as a function of
strip width $\ell$.  The entanglement entropy plotted in this figure and hereafter is the finite part with the UV divergent part subtracted off. The dashed purple and solid blue curves come from the connected configuration, while the solid red one comes from the disconnected configuration. The lowest curve is physically favored compared with others.}
\end{figure}
For such an embedding, following the formula (\ref{law}), the entanglement entropy of the subsystem is given by
\begin{equation}\label{entropy1}
S_\mathcal{A}^{connect}=\frac{4\pi L}{\kappa^2}R\Gamma \int_{r_*}^{\frac{1}{\epsilon}}\frac{r^4\sqrt{g(r)h(r)}e^{-\chi(r)}}{\sqrt{r^6 g(r)h(r)e^{-\chi(r)}-r_*^6 g(r_*)h(r_*)e^{-\chi(r_*)}}}dr=\frac{2\pi L}{\kappa^2}R\Gamma(\frac{1}{\epsilon^2}+S^{con}),
\end{equation}
where the UV  divergent part $1/\epsilon^2$  has been separated from the total entropy and $S^{con}$ is a finite part. The width $\ell$ of the subsystem $\mathcal{A}$ and $r_*$ are connected by the relation
\begin{equation}\label{width1}
\frac{\ell}{2}=\int_{r_*}^{\frac{1}{\epsilon}}dr\frac{dx}{dr}=\int_{r_*}^{\frac{1}{\epsilon}}
\frac{L}{r^2\sqrt{g(r)h(r)(\frac{r^6 g(r)h(r)e^{-\chi(r)}}{r_*^6 g(r_*)h(r_*)e^{-\chi(r_*)}}-1)}}dr.
\end{equation}
 On the other hand, the disconnected configuration consists of two separated surfaces that are located at $x=\pm\frac{\ell}{2}$  and both start from the AdS boundary $r=1/\epsilon$, extend into the bulk until they reach the tip of the soliton solution $r=r_0$. The entanglement entropy for this disconnected configuration is independent of $\ell$, and given by
\begin{equation}\label{entropy2}
S_\mathcal{A}^{disconnect}=\frac{4\pi L}{\kappa^2}R\Gamma \int_{r_0}^{\frac{1}{\epsilon}} r e^{-\frac{\chi(r)}{2}}dr=\frac{2\pi L}{\kappa^2}R\Gamma(\frac{1}{\epsilon^2}+S^{discon}),
\end{equation}
where $S^{discon}$ is the finite part.

 Once  have numerically solved the metric functions, we can now calculate the entanglement entropy. Following the above discussion we must calculate the entropies of the connected configuration~\eqref{entropy1} and the disconnected one~\eqref{entropy2}, respectively. One must take care of the UV divergence  for the calculations of each entropy, but the difference of entropies is insensitive to it.  Therefore, what we are interested in is the finite part of the entanglement entropy, i.e., $S^{discon}$ and $S^{con}$ in our discussion. We find that the entanglement entropy with respect to strip width $\ell$ behaves quite similar for different choice of parameters, i.e., $\alpha$ and $\mu$, which is schematically drawn in Figure~\eqref{entropywidth}. We stress here that the entanglement entropy drawn in Figure~\eqref{entropywidth} and hereafter is the finite part with the UV divergent part subtracted off. \footnote{We omit the superscript used to distinguish the connected embedding and disconnected one for convenience. This will not make any confusion since it is easy to distinguish these two cases from the context.} We find that there exists a maximal width $\ell_{max}$. For $\ell>\ell_{max}$, the connected configuration does not exist and the physical solution comes from the trivial disconnected configuration where the entropy is independent of $\ell$. On the contrary, there are three different branches for the entanglement entropy  when $\ell < \ell_{max}$. Two of them come from connected configurations and the third one corresponds to the disconnected one. The physical entropy can be found by always choosing the lowest branch since it has minimal value. As we can see from Figure~\eqref{entropywidth}, the lower connected branch is favored for sufficiently small $\ell$. Hence, there is a critical value $\ell_{critical}$ below which the connected branch is physically favored. Thus, as we change $\ell$, a phase transition must occur at $\ell_{critical}$, which is just the so called ``confinement/deconfinement" phase transition discussed intensively, for example, in Refs.~\cite{con1,con2,Myers:2012ed}. To be precise, for $\ell<\ell_{critical}$, the entanglement entropy comes from the connected configuration and exhibits non-trivial dependence on $\ell$, which describes a ``deconfining" phase. For $\ell>\ell_{critical}$, the entropy is dominated by the disconnected configuration and is $\ell$ independent, which indicates a ``confining" phase.
\begin{figure}[h]
\centering
\includegraphics[scale=0.92]{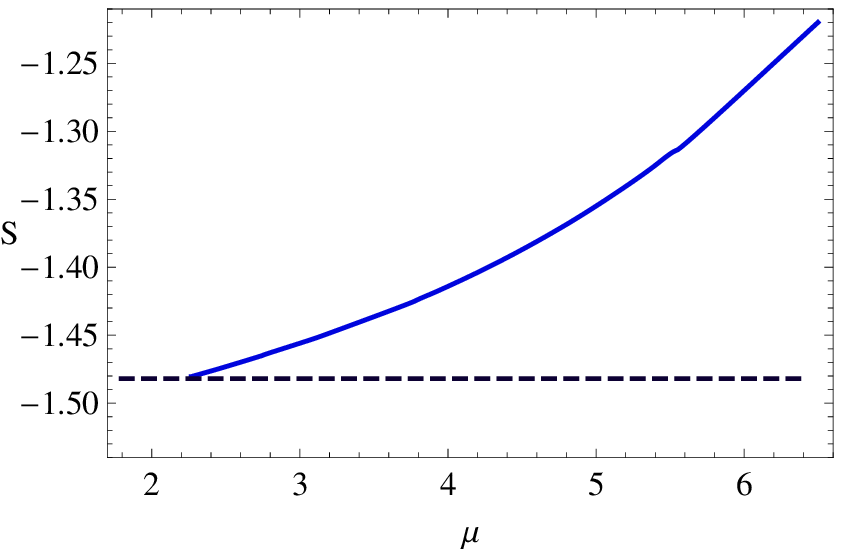}\ \ \ \
\includegraphics[scale=0.9]{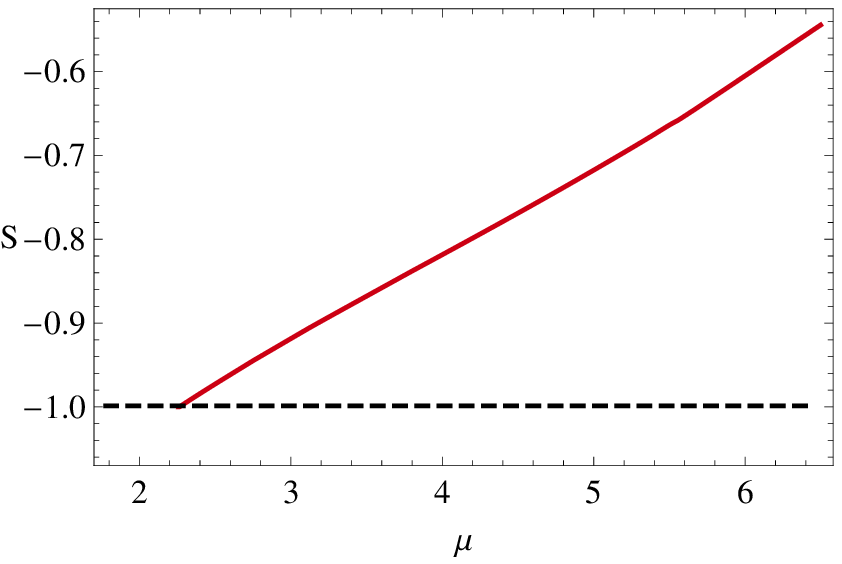} \caption{\label{pentropyxtwo} The entanglement entropy as a function of chemical potential for strip with finite length along $x$ direction. We choose $\alpha=0.2$ at fixed $\ell=0.5L$ (within ``deconfining phase", left plot) and $\ell\longrightarrow\infty$ (within ``confining phase", right plot), respectively. The transition from the insulator to superconductor is second order here. In both plots, the solid curves come from superconducting phase, while the dashed black lines come from insulator phase. The critical chemical potential here is $\mu_c\simeq2.265$. The physical curve is determined by choosing the dashed line below $\mu_c$ and solid  curve above $\mu_c$.}
\end{figure}

Thus we have totally four phases in the boundary field theory side, i.e., the confining/deconfining phases for the insulator and superconductor~\cite{Cai:2012sk}, respectively, which are characterized by the chemical potential $\mu$ and strip width $\ell$. In particular, the strip width controls the ``confinement/deconfinement" phase transition.~\footnote{Strictly speaking, the term phase transition here is inappropriate since the system itself, i.e., the state of the boundary field theory, does not change at all as one changes $\ell$. However, this behavior is reminiscent of many thermodynamic phase transitions in holographic calculations (see, for example Refs.~\cite{con1,con2,Myers:2012ed}) and so here we adopt the terminology ``phase transition" to convey this picture as in the literature.}

\begin{figure}[h]
\centering
\includegraphics[scale=0.92]{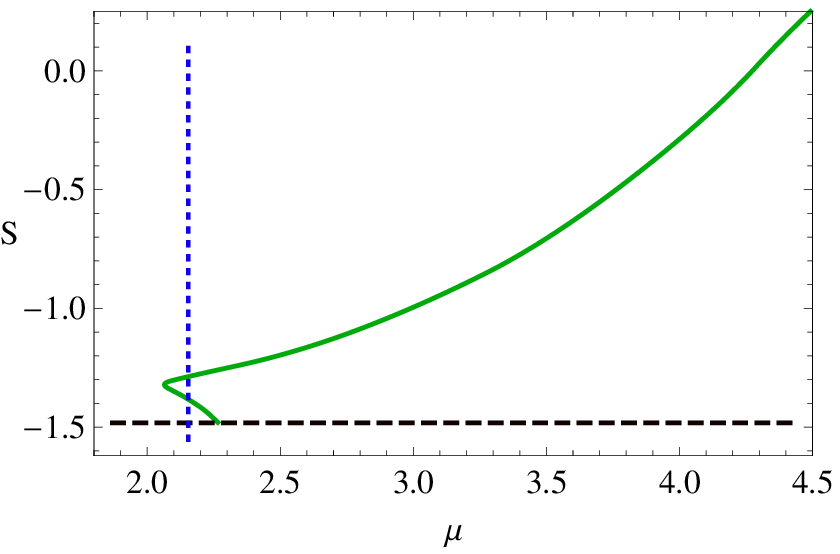}\ \ \ \
\includegraphics[scale=0.9]{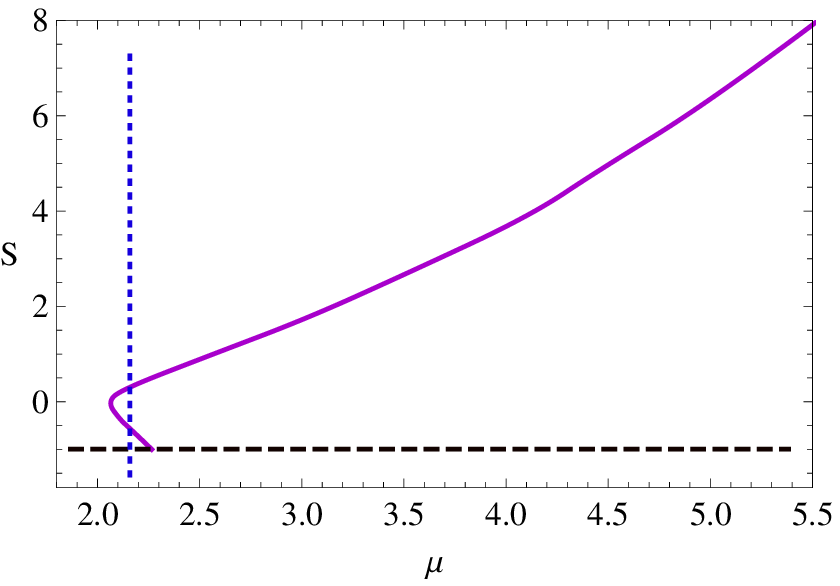} \caption{\label{pentropyxfir} The entanglement entropy as a function of chemical potential for strip with finite length along $x$ direction. We choose $\alpha=0.6$ at fixed $\ell=0.5L$ (within ``deconfining phase", left plot) and $\ell\longrightarrow\infty$ (within ``confining phase", right plot), respectively. The transition from the insulator to superconductor is first order here. The critical chemical potential $\mu_c\simeq2.157$ is indicated by the vertical dotted blue line. In both plots, the solid curves come from the superconducting phase, while the dashed black lines come from the insulator phase. The physical curve is determined by choosing the dashed line below $\mu_c$ and solid upper curve above $\mu_c$.}
\end{figure}

It is instructive to study how the entropy of the subsystem changes with respect to chemical potential at fixed $\ell$. We first focus on the second order transition case, which is presented in Figure~\eqref{pentropyxtwo}. We can see that, from the insulator phase to the superconductor phase, the entanglement entropy increases monotonically  with the increment of chemical potential both in ``deconfining phase" and ``confining phase".~\footnote{Due to the lake of numerical control at large chemical potential, we are not able to give the value of entropy for very large $\mu$. But up to $\mu\simeq15$, we have confirmed that the trend of the entanglement entropy curve does not change. } Furthermore, the entanglement entropy is continuous at critical chemical potential $\mu_c$, but its slope has a discontinuous change at $\mu_c$. Comparing with Figure~\eqref{pentropyxtwo}, we can see a dramatic change in the first order case drawn in Figure~\eqref{pentropyxfir}. Although the entropy increases, there is an sudden jump in entropy as well as its slope at $\mu_c$. The discontinuity or jump at critical point indicates some kind of significant reorganization of the degrees of freedom of the system, since new degrees of freedom are expected to emerge in the new phase.

\begin{figure}[h]
\centering
\includegraphics[scale=0.9]{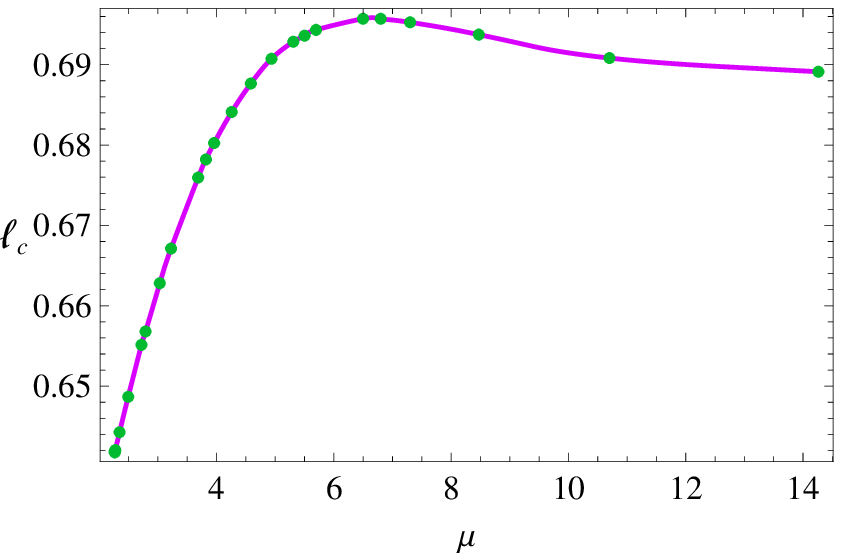}\ \ \ \
\includegraphics[scale=0.9]{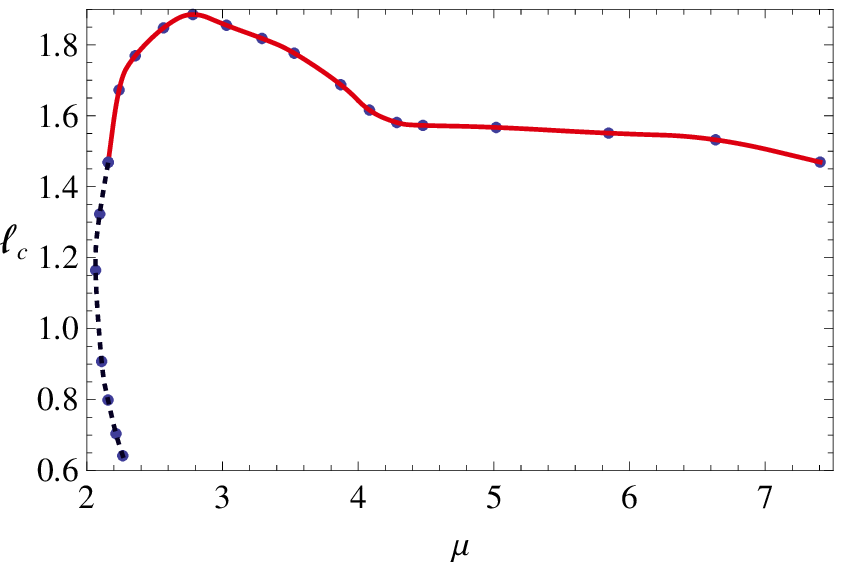} \caption{\label{pcriticax} The critical width as a function of chemical potential in superconducting phase for $\alpha=0.2$ with $\mu_c\simeq2.265$ (left plot) and $\alpha=0.6$ with $\mu_c\simeq2.157$ (right plot), respectively. For the first order phase transition case (right plot), the physical curve denoted by solid curve starts from the critical chemical potential $\mu_c\simeq2.157$. }
\end{figure}

No matter of the order of the transition, the entanglement entropy increases as one increases chemical potential for the p-wave superconductor/insulator case. Such behavior is quite different from the s-wave superconductor/insulator case, where the entropy with respect to chemical potential is non-monotonic in the superconducting phase. However, we can find that the critical width as a function of chemical potential in the superconducting phase behaves non-monotonic, which is presented in Figure~\eqref{pcriticax}. This behavior is qualitatively same as the
 s-wave case.

%

\subsection{Strip with finite width along $y$ direction}
\label{sect:Pgravity}
We now consider a straight belt geometry with a finite width $\ell$ along the $y$ direction. The subsystem $\mathcal{B}$ locates in the slice $r=\frac{1}{\epsilon}$ with the UV cutoff $\epsilon\rightarrow0$, and extends in $x$ and $\eta$ directions. The holographic dual surface $\gamma_B$ extending to the bulk, is defined as a three-dimensional surface
\begin{equation}\label{embed}
t=0,\ \ r=r(y),\ \ -\frac{R}{2}<x<\frac{R}{2}\ (R\rightarrow\infty),\ \ 0\leq\eta\leq\Gamma,
\end{equation}
where $R$ is the regularized length in the $x$ direction. In this case, there also exist two kinds of dual surafces: connected configuration and disconnected configuration as
in the previous subsection. Let us first concentrate on the connected configuration which is smooth at the turning point at $r=r_*$. The dual surface $\gamma_B$ starting from $y=\frac{\ell}{2}$ at $r=\frac{1}{\epsilon}$  extends into the bulk until it reaches the turning point $r=r_*$, then goes back to the AdS boundary $r=\frac{1}{\epsilon}$ at $y=-\frac{\ell}{2}$.

In this case, the corresponding entanglement entropy is given by
\begin{equation}\label{entropy1x}
S_\mathcal{B}^{connect}=\frac{4\pi L}{\kappa^2}R\Gamma \int_{r_*}^{\frac{1}{\epsilon}}\frac{r^4\sqrt{g(r)}h(r)e^{-\chi(r)}}{\sqrt{r^6 g(r)h(r)e^{-\chi(r)}-r_*^6 g(r_*)h(r_*)e^{-\chi(r_*)}}}dr=\frac{2\pi L}{\kappa^2}R\Gamma(\frac{1}{\epsilon^2}+S^{con}),
\end{equation}
where the UV divergent part has been separated form the total entropy and $S^{con}$ is a finite part. The width $\ell$ of the subsystem $\mathcal{B}$ and $r_*$ are related by the equation
\begin{equation}\label{width1x}
\frac{\ell}{2}=\int_{r_*}^{\frac{1}{\epsilon}}dr\frac{dy}{dr}=\int_{r_*}^{\frac{1}{\epsilon}}
\frac{L}{r^2\sqrt{g(r)(\frac{r^6 g(r)h(r)e^{-\chi(r)}}{r_*^6 g(r_*)h(r_*)e^{-\chi(r_*)}}-1)}}dr.
\end{equation}
 On the other hand, the disconnected configuration consists of  two separated surfaces are located at $y=\pm\frac{\ell}{2}$, respectively, and extend into the bulk until
 they reach the tip $r=r_0$,  The entropy  for this disconnected configuration is $\ell$ independent and  given by
\begin{equation}\label{entropy2x}
S_\mathcal{B}^{disconnect}=\frac{4\pi L}{\kappa^2}R\Gamma \int_{r_0}^{\frac{1}{\epsilon}} r \sqrt{h(r)}e^{-\frac{\chi(r)}{2}}dr=\frac{2\pi L}{\kappa^2}R\Gamma(\frac{1}{\epsilon^2}+S^{discon}),
\end{equation}
where once again, $S^{dsicon}$ denotes the finite part.
\begin{figure}[h]
\centering
\includegraphics[scale=0.92]{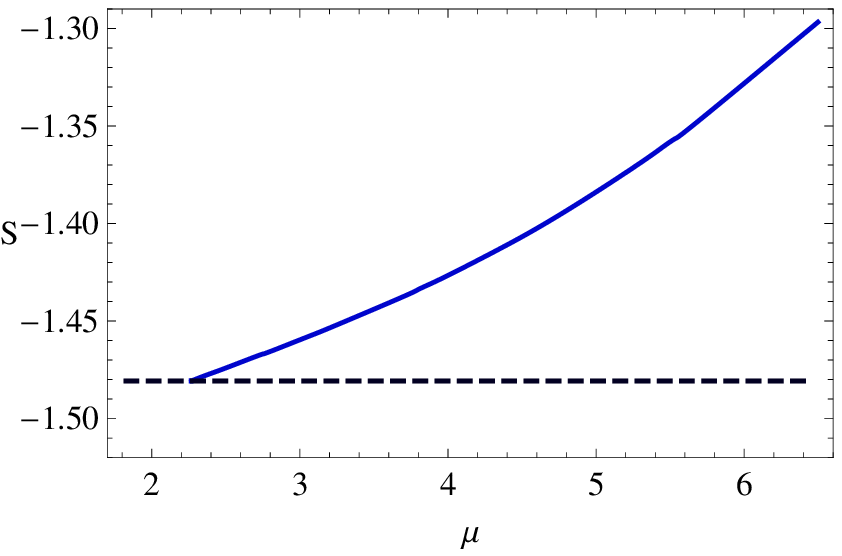}\ \ \ \
\includegraphics[scale=0.9]{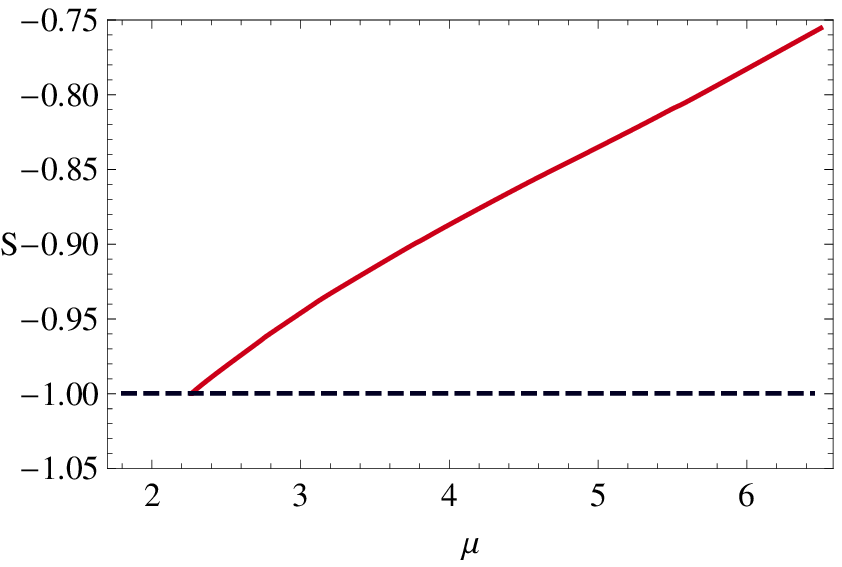} \caption{\label{pentropyytwo} The entanglement entropy as a function of chemical potential for strip with finite length along the $y$ direction. We consider $\alpha=0.2$ at fixed $\ell=0.5L$ (within ``deconfining phase", left plot) and $\ell\longrightarrow\infty$ (within ``confining phase", right plot), respectively. The transition from the insulator to superconductor is second order here with the critical chemical potential $\mu_c\simeq2.265$. In both plots, the solid curves come from the superconducting phase, while the dashed black lines come from the insulator phase. The physical curve is determined by choosing the dashed line below $\mu_c$ and solid upper curve above $\mu_c$.}
\end{figure}
\begin{figure}[h]
\centering
\includegraphics[scale=0.92]{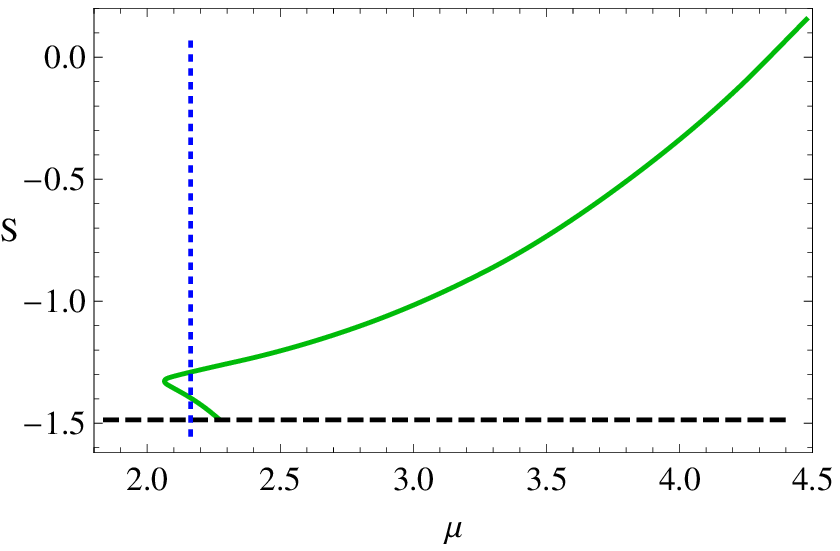}\ \ \ \
\includegraphics[scale=0.9]{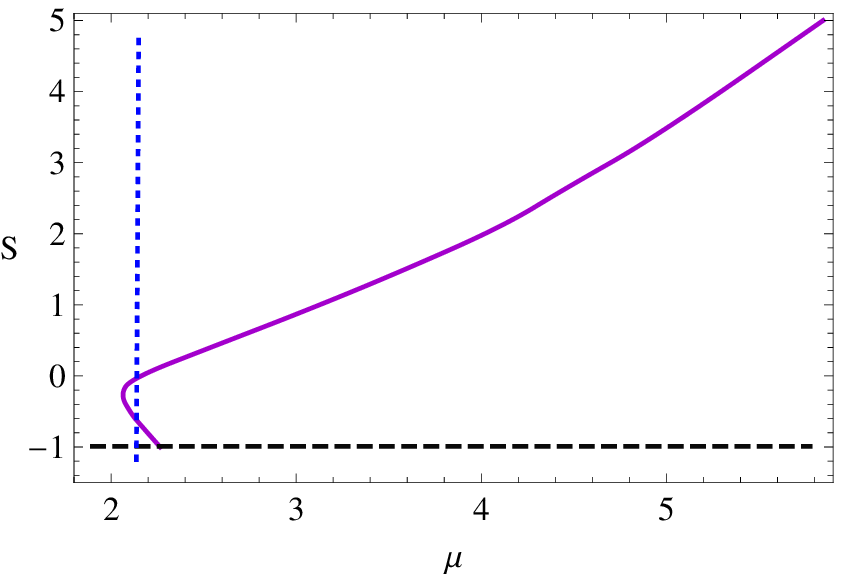} \caption{\label{pentropyyfir} The entanglement entropy as a function of chemical potential for the strip with finite length along the $y$ direction. We consider $\alpha=0.6$ at fixed $\ell=0.5L$ (within ``deconfining phase", left plot) and $\ell\longrightarrow\infty$ (within ``confining phase", right plot), respectively. The transition from the insulator to superconductor is first order here. The critical chemical potential $\mu_c\simeq2.157$ is indicated by the vertical dotted blue line. In both plots, the solid curves come from the superconducting phase, while the dashed black lines come from the insulator phase. The physical curve is determined by choosing the dashed line below $\mu_c$ and solid upper curve above $\mu_c$.}
\end{figure}

\begin{figure}[h]
\centering
\includegraphics[scale=0.9]{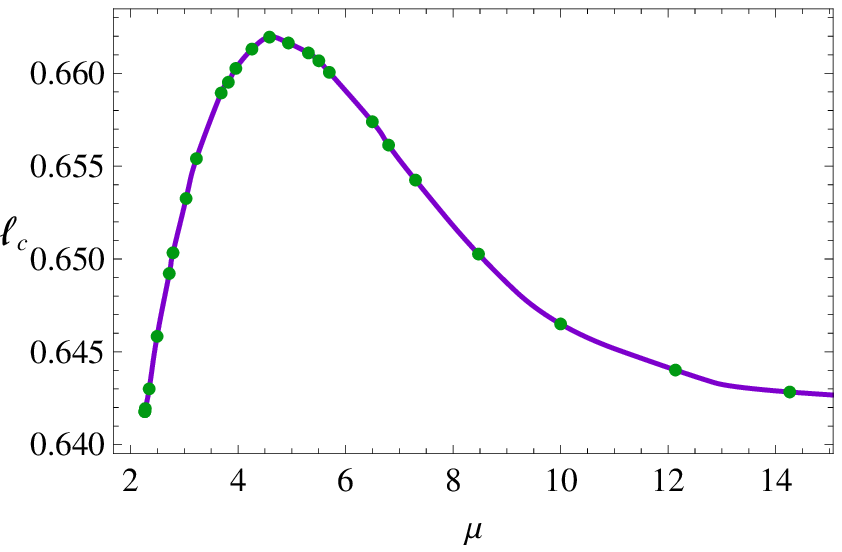}\ \ \ \
\includegraphics[scale=0.9]{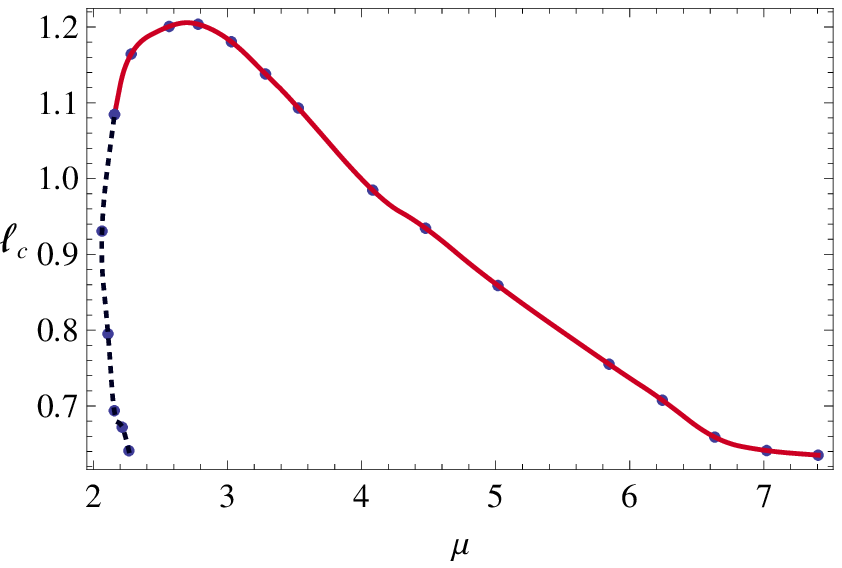} \caption{\label{pcriticay} The critical width as a function of chemical potential in superconducting phase for $\alpha=0.2$ with $\mu_c\simeq2.265$ (left plot) and $\alpha=0.6$ with $\mu_c\simeq2.157$ (right plot), respectively. For the first order phase transition case (right plot), the physical curve denoted by solid curve  starts from the critical chemical potential $\mu_c\simeq2.157$.}
\end{figure}

We find that for the strip along the $y$ direction, the entanglement entropy with respect to strip width $\ell$ behaves similar as the case along the $x$ direction shown in Figure~\eqref{entropywidth} for different choice of parameters, i.e., $\alpha$ and $\mu$. As one tunes strip width, the ``confinement/deconfinement" phase transition always exists. We present the entanglement entropy versus chemical potential in Figure~\eqref{pentropyytwo} for $\alpha=0.2$ and in Figure~\eqref{pentropyyfir} for $\alpha=0.6$, respectively. No matter of the order of the transition, the entanglement entropy increases as one increases chemical potential for the p-wave superconductor/insulator case. The critical width as a function of chemical potential in the superconducting phase also behaves non-monotonic, which is presented in Figure~\eqref{pcriticay}.

Comparing the two setups, i.e., the strip with width $\ell$ along the $x$ direction and the strip with width $\ell$ along the $y$ direction, the results are qualitatively similar. However, the values of entanglement entropy and critical width in the former case are larger for given parameters than in the latter case and the difference grows quickly as one increases the chemical potential. Since the degree of anisotropy of the superconducting phase is characterized by the value of condensate. The growth of the difference of the entanglement entropy in two setups suggests that the entanglement entropy indeed describes the new degrees of freedom emerging in the superconducting phase. As a non-local quantity, the entanglement entropy for different configurations considered in this paper is not expected to have qualitatively difference, which is indeed confirmed by our numerical calculation. On the other hand, if considering local quantities for holographic p-wave model, such as conductivity~\cite{Gubser:2008wv,CNZ1}, one can find that the conductivity $\sigma_{yy}$ orthogonal to the super current $\langle\hat{J}^x_1\rangle$ displays gapped dependence similar to the findings of s-wave case. The conductivity $\sigma_{xx}$ parallel to the super current $\langle\hat{J}^x_1\rangle$ is qualitatively unlike $\sigma_{yy}$ and the low-frequency behavior of $Re[\sigma_{xx}]$ can be characterized very accurately in terms of the Drude model. The essential difference between two conductivities along different directions is the consequence of the anisotropy in the superconducting phase.
%
%
%

\section{Conclusion and discussions}
\label{sect:conclusion}
In our previous papers~\cite{Cai:2012sk,Cai:2012es}, we studied the properties of entanglement entropy in the holographic s-wave superconductor/insulator transitions, where a non-monotonic behavior of the entanglement entropy with respect to chemical potential was found. Such behavior exists for more complex interaction~\cite{Cai:2012es}. In order to see whether the non-monotonic behavior is universal or not, in this paper we generalized our study to the p-wave case. Depending on the strength of the back reaction, the transition between the insulator phase and superconducting phase can be either second order or first order. However, unlike the case in s-wave model, there is no additional first order transition inside the superconducting phase.

With the aim of providing more information on some universal properties of holographic superconductors which may shed toy-model-based insights on understanding the real high $T_c$ superconductors, we continued the study of the behavior of entanglement entropy in the holographic superconducting phase transition. We limited ourselves in this paper to investigate an Einstein-Yang-Mills model, which is dual to the p-wave superconductor/insulator phase transition. In this model, the rotational symmetry acting on $x$ and $y$ is also broken in the superconducting phase. In order to account for such anisotropy, we calculated the entanglement entropy for two kinds of strip geometry with finite width $\ell$ along $x$ direction and $y$ direction, respectively. In both cases, the ``confinement/deconfinement" phase transition always appears. No matter of the order of the phase transition, the entanglement entropy as a function of chemical potential monotonically increases, which is much different from the s-wave case. Since the entanglement entropy is a non-local quantity in the field theory side and measures the area of a minimal surface extending into the bulk in the holographic description, we do not expect that the entanglement entropy for the two kinds of strip geometry will be qualitatively different. Our calculations indeed confirmed this and the entanglement entropy has a difference only quantitatively. More precisely, the difference increases as the condensate increases. Thus, it is reasonable to conclude that the entanglement entropy indeed describes the new degrees of freedom emerging in the superconducting phase. We also extracted the critical width which behaves non-monotonic with respect to chemical potential. This behavior is the same as the one in the s-wave case.  In addition, we further provided the evidence that the entanglement entropy can indeed indicate the appearance and the order of phase transition: In the second order phase transition case, the entanglement entropy is continuous, but its slope has a jump at the critical chemical potential, while in the first order transition case, both the entanglement entropy and its slope have a jump at the critical chemical potential.

The different behavior of entanglement entropy in the s-wave case and p-wave case makes the aim to understand the non-monotonic behavior difficult at the moment. Nevertheless, after a second thought, the different behavior of entanglement entropy in both cases looks not so surprised, since the properties of holographic s-wave model and p-wave model are quite different. Ref.~\cite{Horowitz:2010jq} studied $(3+1)$-dimensional s-wave superconductor/conductor/insulator phase transition. For the conductor/superconductor case, the order of the phase transition is always second order. In contrast to the conductor/superconductor case, the insulator/superconductor phase transition is second order for small back reaction, for intermediate back reaction, there is an additional first order transition inside the superconducting phase and the transition finally becomes first order for sufficiently large back reaction. Similar phenomena exist even when one introduces more complex interaction~\cite{Cai:2012es}. Actually, we have also checked the same model in one dimension less, i.e., in $(2+1)$-dimensional case and found the same result. Let us now consider holographic p-wave models. The full back reacted p-wave superconductor/conductor transition in $(3+1)$ dimensions studied in~Ref.~\cite{Ammon:2009xh} shows that the order of the phase transition changes from second order to first order as one increases the strength of the back reaction, while for the $(2+1)$-dimensional case, the transition is always second order~\cite{Arias:2012py}. The $(3+1)$-dimensional p-wave superconductor/insulator model studied in this paper and Ref.~\cite{Akhavan:2010bf} behaves very similar to the $(3+1)$ dimensional p-wave superconductor/conductor model.  Let us note that the order of the phase transition for the p-wave case depends on not only the dimension of space-time, but also the strength of the back reaction. This is quite different from the s-wave case. Furthermore, let us notice that in Ref.~\cite{2006PhRvB..73x5115R}, the entanglement entropy for a $(2+1)$-dimensional chiral p-wave superconductor model is calculated by dividing the system as two parts separated by a hyperplane. In this model, there are four phases separated by three quantum critical points at chemical potential $\mu=0$ and $\pm 4$, which are labeled by the Chern number  $Ch$ as $Ch=0$ $(|\mu |>4) $, $Ch=-1$ $ (-4 < \mu <0)$, and $Ch =+1$ $ (0 <\mu <+4)$. The entanglement entropy as a function of chemical potential is presented in Figure 4-(e) of Ref.~\cite{2006PhRvB..73x5115R}. One can see from the plot that the behavior of the entanglement entropy depends on not only the phase, but also the aspect ratio $r=N_y/N_x$, the ratio of lattice number in  $y$ direction over in  $x$ direction. In particular,  in some phases
both the monotonic and non-monotonic behaviors appear depending on the value of $r$. For a large $r$, the entanglement entropy is monotonic in a fixed phase while it becomes non-monotonic for a small $r$. It indicates that the entanglement entropy can indeed reveal many properties of many-body systems.

In our setup, the holographic superconductor extends infinitely along $x$ and $y$ directions. In some sense this case corresponds to the large r situation discussed in Ref.\cite{2006PhRvB..73x5115R}. Therefore consider the entanglement entropy as a measure of the degrees of freedom for a given system, the monotonic behavior of the entanglement entropy in the holographic p-wave superconductor/insulator arises naturally. The normal insulator phase can be thought of as having little degrees of freedom due to the mass gap, which is consistent with the vanishing charge density. As we increase the chemical potential, the superconducting phase appears. The growth of the condensate and the charge density  indicates that the total degrees of freedom increase, which just accounts for the increase of the entanglement entropy.  However, it is still obviously necessary to understand the different behavior of entanglement entropy in holographic s-wave and p-wave superconductor/insulator transitions. For this, one needs know the details of dual field theories for these two kinds of models.  In principle, one can finish this task if one could embed these two models into string theory or M-theory.  Unfortunately, we have not yet had the microscopic theories for these two models. Nevertheless, the phenomenological bottom-up approach we adopted here can provide some basic information of a particular class of strongly coupled field theory. In this sense it might also helpful to use the entanglement entropy as a probe to study the holographic d-wave models~\cite{Ge:2012vp}.

\section*{Acknowledgements}     This work was supported in part by the National Natural Science Foundation of China (No.10821504, No.10975168 and No.11035008,No.11205226,No.11175019), and
in part by the Ministry of Science and Technology of China under Grant No. 2010CB833004. LFL would like appreciate the general financial support from China Postdoctoral Science Foundation No. 2012M510563. LL and LFL would like to thank Hai-Qing Zhang, Yun-Long Zhang and Song He for useful discussions. LL is extremely grateful to Yong-Qiang Wang for his help in the numerical calculations.

\appendix

\section{Numerical details}
\label{sect:details}

In this Appendix, we give the details for solving the coupled equations of motion~\eqref{eoms}. In order to find the solutions for all the six functions $\mathcal{F}=\{\phi,\omega,f,h,g,\chi\}$, one must impose suitable boundary conditions at boundary $r\rightarrow\infty$ and  at the tip $r=r_0$. The general falloff near the boundary is given by~\eqref{boundary} where we choose $\omega_0=0$ since one wants the condensate to arise spontaneously.

Regularity of the solution at the tip $r=r_0$ requires that all six functions have finite values at the tip and  have Taylor series expansions near the tip
\begin{equation}\label{series}
\mathcal{F}=\mathcal{F}(r_0)+\mathcal{F}'(r_0)(r-r_0)+\cdots.
\end{equation}
Plugging the expansion~\eqref{series} into~\eqref{eoms}, and using $g(r_0)=0$, we are left with six independent parameters $\{r_0,\phi(r_0),\omega(r_0),f(r_0),h(r_0),\chi(r_0)\}$. However,
the equations of motion~\eqref{eoms} have four useful scaling symmetries
\begin{equation} \label{scaling1}
f\rightarrow \lambda^2 f,\quad \phi\rightarrow\lambda\phi,
\end{equation}

\begin{equation} \label{scaling2}
h\rightarrow\lambda^2 h,\quad \omega\rightarrow\lambda\omega,
\end{equation}

\begin{equation} \label{scaling3}
\chi\rightarrow \chi+\lambda,\quad \eta\rightarrow\eta e^{\lambda/2},
\end{equation}

\begin{equation} \label{scaling4}
r\rightarrow\lambda r,\quad (t,x,y,\eta)\rightarrow{\lambda^{-1}}(t,x,y,\eta),\quad(\phi,\omega)\rightarrow\lambda(\phi,\omega).
\end{equation}
Taking advantage of such four scaling symmetries, we can firstly set $\{r_0=1,f(r_0)=1,h(r_0)=1,\chi(r_0)=0\}$ for performing numerics. After solving the coupled differential equations, we should use the first three symmetries again to satisfy the asymptotic conditions. We will choose $\phi(r_0)$ as a shooting parameter to match the source free condition, i.e, $\omega_0=0$.

All in all, for each choice of $\alpha$ and $\omega(r_0)$, we can solve the equations of motion~\eqref{eoms}. After solving the equations, we can obtain the current condensate $\langle\hat{J}^x_1\rangle=\frac{2\alpha^2}{\kappa^2 L}w_2$, chemical potential $\mu=\phi_0$ and charge density $\rho=\frac{2\alpha^2}{\kappa^2 L}\phi_2$ by reading off the coefficients $w_2$, $\phi_0$ and $\phi_2$ in~\eqref{boundary}, respectively.

\begin{figure}[h]
\centering
\includegraphics[scale=0.74]{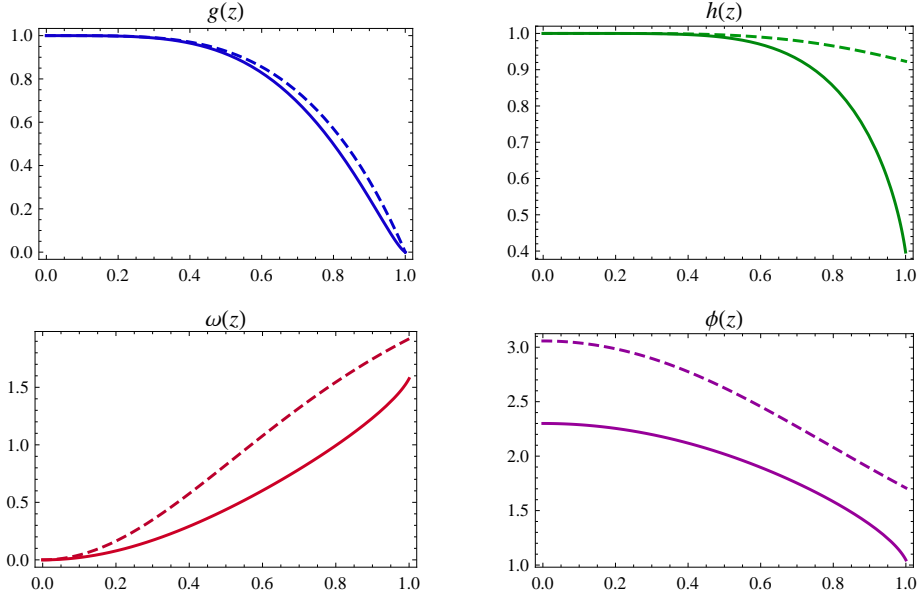}\caption{\label{pfunction}  Typical soliton solution with nonvanishing vector hair. The solid curves stand for the case with $\alpha=0.6$, where the value of $\omega$ at tip is $\omega(z_0)\simeq 1.576 $ and the corresponding identification length $\Gamma\simeq2.842$, while the dashed lines are for the case with $\alpha=0.2$, where the value of $\omega$ at tip $\omega(z_0) \sim 1.922$ and the corresponding identification length $\Gamma \sim 3.118$.}
\end{figure}

In fact, it is convenient in the numerical calculations to make a coordinate transformation from $r$ coordinate to $z$ coordinate by defining $z = 1/r$.  In that case, the infinite boundary is now at $z=0$ and the tip
sits at $z_0 =1/r_0=1$. Two typical sets of solution for the metric and gauge field configurations are presented in Figure~\eqref{pfunction} for $\alpha=0.2$ and
$\alpha=0.6$. respectively. Note that the metric functions $g$ and $h$ will be used in calculating the holographic entanglement entropy.
For fixed strength of back reaction $\alpha$, one has a one parameter family of solutions denoted by the value of $\omega$ at the tip. However, for different choices of $\omega(r_0)$, the identification length $\Gamma$ in $\eta$ coordinate will be different. In order to compare different solutions appropriately, the boundary geometry should be the same. Making use of the scaling symmetry~\eqref{scaling4}, one can find that the relevant quantities scale as follows
\begin{equation} \label{scalingvalue}
\begin{split}
&\Gamma\rightarrow \Gamma/\lambda,\quad r_0\rightarrow \lambda r_0,\quad \mu\rightarrow\lambda\mu,\quad \rho\rightarrow\lambda^2\rho,\quad \langle\hat{J}^x_1\rangle\rightarrow\lambda^2\langle\hat{J}^x_1\rangle,\\
&S^{con/discon}\rightarrow\lambda^2S^{con/discon},\quad \ell\rightarrow\ell/\lambda,
\end{split}
\end{equation}
where $S^{con/discon}$  stands for the finite part of the entanglement entropy defined in section~\eqref{sect:Phee} and $\ell$ is the strip width. Since in our units setup, the identification
length $\Gamma$ in the pure soliton is $\pi$, we will scale all $\Gamma$ for each solution to be $\pi$. It should be stressed  here that after the scaling transformation, the tip $r_0$ will be no longer at $r_0=1$.
\begin{figure}[h]
\centering
\includegraphics[scale=1.2]{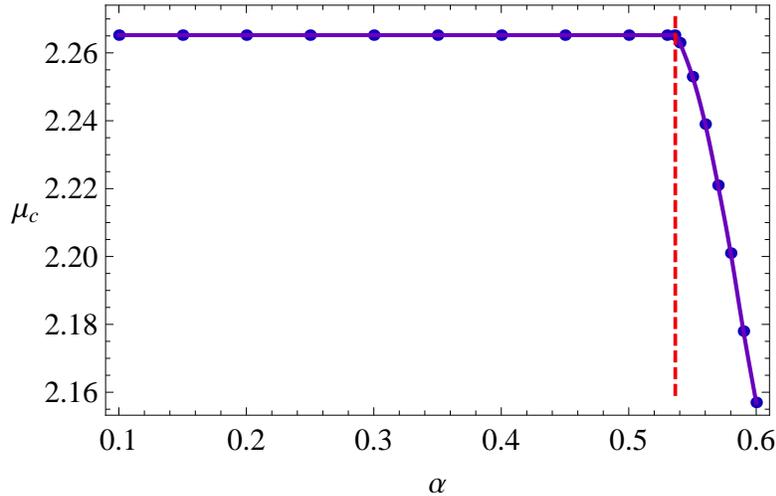}\caption{\label{critmu} The critical chemical potential $\mu_c$ as a function of the strength of back reaction $\alpha$. The critical strength $\alpha_c\simeq0.538\pm0.002$, denoted by the dashed vertical line, where the order of transition is changed. Here we scale all $\Gamma$ for each $\alpha$ to be $\pi$.}
\end{figure}

For each $\alpha$, as one increases the chemical potential, the solution with non-vanishing vector ``hair" will appear and will be finally thermodynamically favored over the pure AdS soliton solution above the critical chemical potential $\mu_c$, where a phase transition from insulator phase to superconducting phase happens. We present critical chemical potential $\mu_c$ as a function of the strength of the back reaction $\alpha$ in Figure~\eqref{critmu}. For small $\alpha$, the transition is second order and $\mu_c$ is almost independent of the strength of the back reaction. For larger $\alpha$, the transition becomes first order and the critical chemical potential decreases as  $\alpha$ increases. Therefore, there exists a critical value $\alpha_c\simeq0.538\pm0.002$, beyond which the phase transition changes from  second order to  first order.

\end{document}